\documentclass[aip,graphicx]{revtex4-1}
\usepackage{amsmath}
\usepackage{epsfig}
\usepackage{graphicx}
\usepackage{dcolumn}
\usepackage{bm}

\begin{document}

\title{{\Large Electromagnetic Scattering by Spheres of Topological Insulators}}

\author{Lixin Ge}
\affiliation{Department of Applied Physics, College of Physics,
Chongqing University, Chongqing 400044, China}
\affiliation{Department of Physics, Key laboratory of Micro and Nano Photonic Structures
(Ministry of Education), and Key Laboratory of Surface Physics, Fudan
University, Shanghai 200433, P. R. China}

\author{Dezhuan Han}
\email{dzhan@cqu.edu.cn}
\affiliation{Department of Applied Physics, College of Physics,
Chongqing University, Chongqing 400044, China}

\author{Jian Zi}
\email{jzi@fudan.edu.cn}
\affiliation{Department of Physics, Key laboratory of Micro and Nano Photonic Structures
(Ministry of Education), and Key Laboratory of Surface Physics, Fudan
University, Shanghai 200433, P. R. China}
\date{\today }

\begin{abstract}
The electromagnetic scattering properties of topological insulator (TI) spheres are systematically studied in this paper. Unconventional backward scattering caused by the topological magneto-electric (TME) effect of TIs are found in both Rayleigh and Mie scattering regimes. This enhanced backward scattering can be achieved by introducing an impedance-matched background which can suppress the bulk scattering. For the cross-polarized scattering coefficients, interesting antiresonances are found in the Mie scattering regime, wherein the cross-polarized electromagnetic fields induced by the TME effect are trapped inside TI spheres. In the Rayleigh limit, the quantized TME effect of TIs can be determined by measuring the electric-field components of scattered waves in the far field.
\end{abstract}

\maketitle

\section{Introduction}

\bigskip Topological insulators (TIs) are an emerging quantum phase in condensed
matter physics \cite{qi11,has10,qi08}. The key feature of TIs is that the electronic band structure resembles an ordinary insulator in the bulk, while gapless edge or
surface states exist within the bulk energy gap \cite{qi08}. These edge
states are protected by time-reversal symmetry. TI materials have been
theoretically predicted and experimentally observed in various systems such
as HgTe/CdTe and InAs/GaSb quantum well, and Bi$_{1-x}$Sb$_{x}$, Bi$_{2} $Te$%
_{3}$, Bi$_{2}$Se$_{3}$, Bi$_{2}$Te$_{2}$Se \cite{qi11, has10}. A novel quantized
topological magnetoelectric effect (TME) is predicted in TIs that an applied
electric field could induce parallel magnetization and an applied magnetic
field could induce parallel electric polarization \cite{qi08}. An additional
term in the Lagrangian: $\Delta{\mathcal{L}}=(\Theta\alpha/4\pi^{2})\mathbf{E}\cdot%
\mathbf{B}$, which gives rise to the ``axion electrodynamics", contains a
complete description of the electromagnetic(EM) responses of TIs \cite{qi08,wil87}. Here $\alpha=e^{2}/\hbar c$ is the fine structure constant, $\Theta =(2p+1)\pi$ is the
axion angle with $p$ being an integer, $\mathbf{E}$ is electric field and $%
\mathbf{B}$ is magnetic field. Owing to this additional Lagrangian, the constitutive
relations for TIs should be modified as: $\mathbf{D}=\mathbf{\varepsilon E}-
\overline{\alpha} \mathbf{B}, \mathbf{H}=\mathbf{B}/\mu +\overline{\alpha}
\mathbf{E}$, where $\mathbf{D}$ and $\mathbf{H}$ are respectively the
electric displacement and magnetic field strength, $\varepsilon$ and $\mu$
are respectively the permittivity and permeability of TIs, $\overline{\alpha}%
=\Theta \alpha/\pi $ is proportional to the fine structure constant. These modified constitutive relations provide us a compact form to deal with the TME effect as the doping level is in the gap. Many unusual phenomena due to TME effect have been revealed  \cite{qi09,tse10,mac10}.

The study of EM waves scattering by small particles and its
application is an enormous field \cite{boh83}. EM waves scattering by scatterers of TIs exhibits lots of interesting properties due to the TME effect. For instance, parity-violating light scattering under oblique incidence, and strong perturbation of dipole radiation near the TI spheres are demonstrated \cite{och12}. Broadband strong scattering in backward direction and interesting antiresonances are found for TI cylinders \cite{ge14a}. The antiresonance effect reveals that the TME effect is a surface effect essentially in the language of EM scattering.
Moreover, the quantization of TME effect of TI cylinders can even be determined by
measuring the electric-field components of scattered waves in the far fields
at one or two scattering angles in the Rayleigh scattering limit \cite{ge14b}.

In this article, we study light scattering by spheres of TIs. Based on Mie theory, we derive the scattering coefficients and scattering matrix for TI spheres analytically. Exotic strong backward scattering due to TME effect are also found in both Rayleigh and Mie scattering regimes, similar to the case for TI cylinders. At certain frequencies, antiresonances of cross-polarized scattering coefficients of TI spheres are revealed, wherein the cross-polarized fields induced by TME effect are trapped inside TI spheres. In the Rayleigh limit, we propose a simple way to determine the quantized TME
effect of TIs by measuring the electric-field components of scattered waves in the far field.

\section{Mie theory for TI spheres}

\begin{figure}[htbp]
\centering\includegraphics[width=3cm]{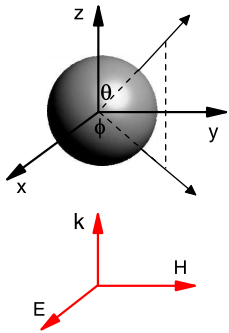}
\caption{Schematic view of the system under study. A TI sphere is placed at the origin. An EM plane wave is incident along $z$ direction and polarized along $x$ direction.}
\label{fig1}
\end{figure}
\bigskip The system under study is shown in Fig. \ref{fig1}. We consider a TI sphere which is illuminated by a time-harmonic EM wave with frequency $\omega$. The radius of the sphere is $a$. The dielectric permittivity and magnetic permeability for the TI sphere are ($\varepsilon_1, \mu_1$), and those for the background medium are ($\varepsilon_b, \mu_b$). This scattering problem can be solved analytically by the standard Mie theory. In the spherical coordinate system, ($r$, $\theta$, $\phi$),
the EM fields can be expanded by the vector spherical harmonics \cite{boh83}:
$\mathbf{M}_{e \mathrm{1} n}^{(I)}(kr)$, $\mathbf{M}_{o\mathrm{1} n}^{(I)}(kr)$, $\mathbf{N}_{e\mathrm{1} n}^{(I)}(kr)$ and $\mathbf{N}_{o\mathrm{1} n}^{(I)}(kr)$, where subscripts $e$ and $o$ denote respectively the even and odd modes with respect to the $x$ axis, $k$ is the corresponding wavevector ($k_{1}=\sqrt{\varepsilon _{1}\mu_{1}}\omega /c$, $k_{b}=\sqrt{\varepsilon_{b}\mu _{b}}\omega /c$ are the wavevectors in the TI sphere and background respectively), $n$ is a integer, and $I$ stands for which kind of spherical Bessel(Hankel) functions we use. We assume that the incident EM wave is $\mathbf{E}_{inc}=E_0 e^{ikz}\hat{e}_x$, where $E_0$ is the amplitude. The scattered and internal electric fields can be expanded in terms of the vector spherical harmonics \cite{boh83}:

\begin{eqnarray}
\mathbf{E}_{sca} &=& \sum^{n=\infty} _{n=1} E_{n}(-b_{n}\mathbf{M}_{o1n}^{(3)}+ia_{n}\mathbf{N}_{e1n}^{(3)}-b_{n}^{TI}%
\mathbf{M}_{e1n}^{(3)}+ia_{n}^{TI}\mathbf{N}_{o1n}^{(3)}) \\
\mathbf{E}_{int} &=& \sum^{n=\infty} _{n=1}E_{n}(c_{n}%
\mathbf{M}_{o1n}^{(1)}-id_{n}\mathbf{N}%
_{e1n}^{(1)}+c_{n}^{TI}\mathbf{M}_{e1n}^{(1)}-id_{n}^{TI}%
\mathbf{N}_{o1n}^{(1)})
\end{eqnarray}
where $E_{n}=i^{n}E_{0}(2n+1)/n(n+1)$, \{$a_{n}, b_{n}$\} are the scattering coefficients that have one-to-one correspondence to the ones for the conventional dielectrics. However, two new scattering coefficients, \{$a_{n}^{TI}$, $b_{n}^{TI}$\}, or the so called \textit{cross-polarized} scattering coefficients arise from the TME effect. The superscripts 1 and 3 stand for the spherical Bessel function of the first kind and spherical Hankel function of the first kind, respectively. By matching the standard boundary conditions at $r=a $, and applying the modified constitutive relations $\mathbf{D} =%
\mathbf{\varepsilon E}- \overline{\alpha} \mathbf{B}, \mathbf{H} =\mathbf{B}%
/\mu +\overline{\alpha} \mathbf{E}$ inside the TI sphere, the scattering coefficients $a_{n}$, $b_{n}$, and the internal coefficients $c_{n}$, $d_{n}$ are found to be:
\begin{eqnarray}
a_{n} &=&\frac{\mu _{b}m^{2}j_{n}(mx)[xj_{n}(x)]^{^{\prime }}\beta _{2}-\mu
_{1}j_{n}(x)[mxj_{n}(mx)]^{^{\prime }}}{\mu
_{b}m^{2}j_{n}(mx)[xh_{n}(x)]^{^{\prime }}\beta _{2}-\mu
_{1}h_{n}(x)[mxj_{n}(mx)]^{^{\prime }}} \\
b_{n} &=&\frac{\mu _{1}j_{n}(mx)[xj_{n}(x)]^{^{\prime }}-\mu
_{b}j_{n}(x)[mxj_{n}(mx)]^{^{\prime }}\beta _{1}}{\mu
_{1}j_{n}(mx)[xh_{n}(x)]^{^{\prime }}-\mu
_{b}h_{n}(x)[mxj_{n}(mx)]^{^{\prime }}\beta _{1}} \\
c_{n} &=&\frac{\mu _{1}j_{n}(x)[xh_{n}(x)]^{^{\prime }}-\mu
_{1}h_{n}(x)[xj_{n}(x)]^{^{\prime }}}{\mu _{1}j_{n}(mx)[xh_{n}(x)]^{^{\prime
}}-\mu _{b}h_{n}(x)[mxj_{n}(mx)]^{^{\prime }}\beta _{1}} \\
d_{n} &=&\frac{\mu _{1}mj_{n}(x)[xh_{n}(x)]^{^{\prime }}-\mu
_{1}mh_{n}(x)[xj_{n}(x)]^{^{\prime }}}{\mu
_{b}m^{2}j_{n}(mx)[xh_{n}(x)]^{^{\prime }}\beta _{2}-\mu
_{1}h_{n}(x)[mxj_{n}(mx)]^{^{\prime }}}
\end{eqnarray}
where $j_n$ and $h_n$ are respectively the spherical Bessel and Hankel functions of the first kind, $x\equiv k_{b}a$ is the size parameter and $m=\sqrt{\varepsilon _{1}\mu
_{1}}/\sqrt{\varepsilon _{b}\mu _{b}}$ is the relative refractive index; $\beta _{1}$ and $\beta _{2}$\ are auxilliary functions related to the axion angle, given by $\beta _{1} =1+ \tilde{\alpha}^{2} \chi_{1}$ and $\beta _{2}=1-\tilde{\alpha}^{2}\chi_{2} $,
where $\tilde{\alpha}=\overline{\alpha }\sqrt{\mu _{1}/\varepsilon _{1}}$ and

\begin{eqnarray}
\chi_{1} &=&\frac{\mu _{b}m^{2}j_{n}(mx)[xh_{n}(x)]^{^{\prime }}}{\mu
_{b}m^{2}j_{n}(mx)[xh_{n}(x)]^{^{\prime }}-\mu
_{1}h_{n}(x)[mxj_{n}(mx)]^{^{\prime }}} \\
\chi_{2} &=&\frac{\mu _{b}h_{n}(x)[mxj_{n}(mx)]^{^{\prime }}}{\mu
_{1}j_{n}(mx)[xh_{n}(x)]^{^{\prime }}-\mu
_{b}h_{n}(x)[mxj_{n}(mx)]^{^{\prime }}}
\end{eqnarray}
The cross-polarized scattering coefficients $a_n^{TI}$ and $b_n^{TI}$,  corresponding to $n$-th order of electric and magnetic multipoles but with $90^\mathrm{o}$ polarization-rotation compared to conventional ones, are given by:
\begin{equation}
a_{n}^{TI} =-\frac{[mxj_{n}(mx)]^{^{\prime }}}{m[xh_{n}(x)]^{^{\prime }}}d_{n}^{TI},~~~ b_{n}^{TI} =-\frac{j_{n}(mx)}{h_{n}(x)}c_{n}^{TI},
\end{equation}
where $d_{n}^{TI}$ and $c_{n}^{TI}$ are the cross-polarized internal coefficients of the TI sphere, given by:
\begin{equation}
 d_{n}^{TI}=\tilde{\alpha} \chi_{1}c_{n},~~~ c_{n}^{TI}=\tilde{\alpha} \chi_{2}d_{n}.
\end{equation}
Obviously, cross-polarized multipolar terms in the scattering fields: $a_n^{TI}$ (electric) and $b_n^{TI}$ (magnetic) are caused directly by the internal multipolar terms: $c_n$ (magnetic) and $d_n$ (electric), respectively. This is exactly a strong manifestation of the TME effect that an applied electric(magnetic) field can induce magnetic(electric) field.
For a topologically trivial insulator, the axion angle $\Theta =0$, namely, $
\tilde{\alpha}=0$, therefore $\beta _{1}=\beta _{2}=1$, $a_n^{TI}=b_n^{TI}=c_n^{TI}=d_n^{TI}=0$, the scattering coefficients $a_n, b_n$ and internal coefficients $c_n, d_n$ are reduced to be the conventional ones \cite{boh83}.

\section{Scattering coefficients in Rayleigh limit}

We assume that $\mu _{1}=\mu _{b}=1$ in this article since the materials considered here are non-magnetic. In the Rayleigh scattering limit ($x\ll 1$ and $mx\ll 1$), it can be shown that only the following scattering coefficients are in the order of $x^{3}$:
\begin{eqnarray}
a_{1} &=&-i\frac{2}{3}\frac{3m^{2}-3+2\overline{\alpha }^{2}/\varepsilon _{b}%
}{3m^{2}+6+2\overline{\alpha }^{2}/\varepsilon _{b}}x^{3}+O(x^{5}) \\
b_{1} &=&i\frac{2}{3}\frac{\overline{\alpha }^{2}/\varepsilon _{b}}{%
3m^{2}+6+2\overline{\alpha }^{2}/\varepsilon _{b}}x^{3}+O(x^{5}) \\
a_{1}^{TI} &=&b_{1}^{TI}=i\frac{2\overline{\alpha }/\sqrt{\varepsilon _{b}}}{%
3m^{2}+6+2\overline{\alpha }^{2}/\varepsilon _{b}}x^{3}+O(x^{5})
\end{eqnarray}
All other scattering coefficients are in the order of $x^{5}$ or higher, and
can be hence neglected in the Rayleigh scattering limit. The scattering coefficients $a_1$ and $b_1$ correspond to the electric and magnetic dipolar terms, whereas $a_{1}^{TI}$  and $b_{1}^{TI}$ correspond to the cross-polarized electric dipolar term and cross-polarized magnetic dipolar term respectively.

\section{Unusual backward scattering}

Since the wavevector of the incident wave is along the $z$ direction, the scattering direction $\hat{\mathbf{e}}_{r}$ and the forward direction $\hat{\mathbf{e}}_{z}$ define a scattering plane. The incident electric fields and the scattered electric fields in the far field ($kr\gg 1$) can be related by a scattering matrix \cite{boh83}:
\begin{equation}
\left(
\begin{array}{cc}
E_{s\parallel }\\
E_{s\perp }%
\end{array}%
\right)=\frac{i\exp (ikr)}{kr}\left(
\begin{array}{cc}
S_{2} & S_{3} \\
S_{4} & S_{1}%
\end{array}%
\right) \left(
\begin{array}{cc}
E_{i\parallel }\\
E_{i\perp }%
\end{array}%
\right)
\end{equation}%
where $E_{s\parallel }$ and $E_{s\perp}$ are respectively the components of
scattered electric fields parallel and perpendicular to the scattering plane.
The corresponding unit vectors parallel and perpendicular to the scattering plane are defined as: $\hat{\mathbf{e}}_{s\parallel }$=$\hat{\mathbf{e}}_{\theta }$ and $\hat{\mathbf{e}}_{s\perp }$=-$\hat{\mathbf{e}}_{\phi }$. The components of incident electric fields parallel and perpendicular to the scattering plane are respectively $E_{i\parallel }=E_{0}\cos \phi $ and $E_{i\perp }=E_{0}\sin \phi $, which depend on azimuth angle. The elements of amplitude scattering matrix $S_{j}$ ($j$%
=1, 2, 3, 4) for a single TI sphere are given by the following relations:
$S_{1} = \sum^{n=\infty} _{n=1} \frac{2n+1}{n(n+1)}\left(a_{n}\pi _{n}+b_{n}\tau _{n}\right)$,
$S_{2} =\sum^{n=\infty} _{n=1}\frac{2n+1}{n(n+1)}\left(a_{n}\tau _{n}+b_{n}\pi _{n}\right)$,
$S_{3} =\sum^{n=\infty} _{n=1}\frac{2n+1}{n(n+1)}\left(a_{n}^{TI}\tau _{n}-b_{n}^{TI}\pi _{n}\right)$,
$S_{4} =\sum^{n=\infty} _{n=1}\frac{2n+1}{n(n+1)}\left(b_{n}^{TI}\tau _{n}-a_{n}^{TI}\pi _{n}\right)$,
where $\pi_n=P_{n}^{1}/\sin\theta$ and $\tau_n=dP_{n}^{1}/d\theta$ \cite{boh83}. $P_{n}^{1}$ are associated Legendre functions.
For conventional dielectric spheres, the condition $a_{n}^{TI}=b_{n}^{TI}=0$
leads to a diagonal scattering matrix, i.e., $S_{3}=S_{4}=0$. For TI
spheres, however, scattering matrix is not a diagonal matrix in general since $%
a_{n}^{TI}$ and $b_{n}^{TI}$ may not vanish owing to the TME effect.

The intensities of scattered waves, defined as energy flux per unit area, characterize scattering properties of the system. Based on the scattering matrix shown above, we'll discuss the role that the dielectric background plays in the EM scattering. In the Rayleigh limit, only electric and magnetic dipolar terms dominate the EM scattering. The angle-dependent functions,  $\pi _{n}$ and $\tau _{n}$ for $n=1$, are given by: $\pi _{1}=1,\tau _{1}=\cos\theta $. Two different kinds of limit can be estimated in the following:

(i), $\epsilon_1-\epsilon_b\gg \bar{\alpha}$. Therefore $|a_{1}|\gg|a_{1}^{TI}|=|b_{1}^{TI}|\gg |b_{1}|$, in other words, the contribution of the conventional electric dipole is much larger than those of the cross-polarized electric and magnetic dipoles induced by TME effect. The intensity of scattered waves is mainly determined by the diagonal terms of scattering matrix. Thus, the scattering patterns of TI spheres are similar to those of conventional dielectric spheres.

(ii), $\epsilon_1-\epsilon_b\ll \bar{\alpha}$. The refractive index of the background is very close to that of the TI sphere. In this case we have $|a_{1}^{TI}|=|b_{1}^{TI}|\gg |a_{1}|,|b_{1}|
$, namely, the conventional electric and magnetic dipoles are negligible
compared to cross-polarized ones. We thus have:
\begin{eqnarray}
I_{s\parallel} \simeq \frac{1}{(kr)^{2}}\left\vert S_{3}\right\vert ^{2}\sin ^{2}\phi =\frac{9}{4}\frac{1}{(kr)^{2}}\left\vert
a_{1}^{TI}\cos \theta -b_{1}^{TI}\right\vert ^{2}\sin ^{2}\phi \\
I_{s\perp} \simeq \frac{1}{(kr)^{2}}\left\vert S_{4}\right\vert ^{2}\cos ^{2}\phi =\frac{9}{4}\frac{1}{(kr)^{2}}\left\vert
b_{1}^{TI}\cos \theta -a_{1}^{TI}\right\vert ^{2}\cos ^{2}\phi.
\end{eqnarray}
where we assume that the incident plane wave has an unit intensity, $I_{s\parallel }$ and $I_{s\perp }$ are intensities correspond to the polarized components with
scattered electric-field vectors parallel and perpendicular to the
scattering plane, respectively.
The scattered intensity is contributed mainly by off-diagonal terms of
scattering matrix. Clearly, in the forward direction ($\theta =0$), the
scattered intensity reduce to zero for both $I_{s\parallel }$ and $%
I_{s\perp }$ since $a_{1}^{TI}=b_{1}^{TI}$. In the backward direction ($%
\theta =\pi $), however, the scattered intensity shows a strong backscattering stemming from the constructive interference of the cross-polarized electric and magnetic dipole radiation, i.e. the terms proportional to $a_{1}^{TI}$ and $b_{1}^{TI}$.
\begin{figure}
\centering\includegraphics[width=7cm]{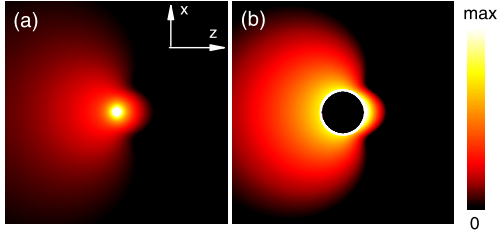}
\caption{Distribution of scattered electric fields $|\mathbf{E}_{sca}|^2$ for TI spheres. The dielectric permittivity of the background medium is the same as that of TI spheres. (a) for Rayleigh scattering ($x$=0.01) and (b) for Mie scattering ($x$=1). The boundary of the sphere in (b) is indicated by a white circle. The incident wave propagates along $z$ direction.}
\label{fig2}
\end{figure}

To show these unusual scattering patterns, we plot the distributions of
scattered electric fields in Fig. \ref{fig2}. The dielectric permittivity of TI sphere and background medium are chosen to be
$\varepsilon_1=\varepsilon_b=30$ .  The resulting scattering patterns are completely different from those for the conventional media, showing a strong backward scattering for both Rayleigh and Mie scattering. It is noted that such enhanced or even complete backward scattering has also been found in other systems such as magnetic\cite{ker83, meh06}, metallic \cite{luk07}, or semiconducting \cite{fu13, gef12} particles. However, unlike metallic spheres \cite{luk07} or semiconducting spheres with high refractive
index \cite{fu13, gef12}, this enhanced backward scattering is a broadband effect for TI spheres since the TME effect is non-resonant intrinsically.

\begin{figure}
\centering\includegraphics[width=7cm]{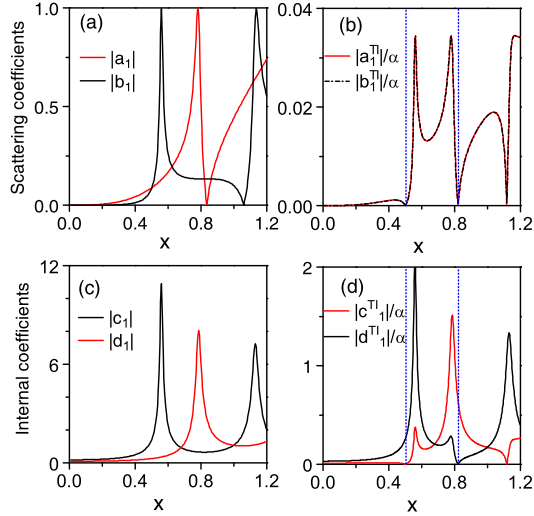}
\caption{Scattering and internal coefficients as functions of the size parameter $x$ with $n=1$. The resonant peaks in (a) and (c) mainly come from the bulk scattering of TI spheres. The corresponding size parameters of antiresonances for cross-polarized coefficients are shown by dotted lines in (b) and (d). The dielectric permittivity of TI spheres is $\epsilon =30$ and the axion angle is chosen to be $\pi$. The background is vacuum.}
\label{fig3}
\end{figure}

\section{Antiresonances}

In Fig. \ref{fig3}, the scattering and internal coefficients corresponding to dipolar modes as functions of size parameter $x$ are shown. For scattering coefficients $a_1$ and $b_1$, there exist resonances and antiresonances. For instance, the resonances at $x=0.780$ for $|a_1|$ and $x=0.560$ for $|b_1|$ correspond to the excitation of electric and magnetic dipolar modes respectively. Meanwhile, the internal coefficients $d_1$ and $c_1$ have similar resonant behaviors as $a_1$ and $b_1$ respectively, which indicates that the EM fields inside the sphere have also been strengthened at the corresponding frequencies. In contrast, at the antiresonances ($x=0.835$ for $|a_1|$ or $x=1.06$ for $|b_1|$) no such electric or magnetic dipolar radiation can be detected outside the sphere since the corresponding scattering coefficients vanish at these particular $x$. However, the electric and magnetic dipolar modes still exist inside the sphere, since the internal coefficients $d_1$ and $c_1$ are actually finite instead of zero as shown in Fig. \ref{fig3}(c).

The scattering coefficients $a_n$ and $b_n$ and internal coefficients $c_n$ and $d_n$ for a TI sphere should be similar to those for a dielectric sphere with the same refractive index, since the fine structure constant is in the order of $10^{-2}$. As a result, TME effect could not be manifested in the scattering coefficients $a_n$, $b_n$ and internal coefficients $c_n, d_n$. However, it plays a key role in the cross-polarized coefficients $a_n^{TI}, b_n^{TI}$ and $c_n^{TI}, d_n^{TI}$ since these four coefficients should vanish in conventional dielectric spheres. Interestingly, there also exist resonances and \textit{antiresonances} for $a_n^{TI}$ and $b_n^{TI}$. The cross-polarized scattering coefficients vanish at the antiresonances. In order to guarantee  $a_n^{TI}=b_n^{TI}=0$, the condition $j_{n}(mx)[mxj_{n}(mx)]^{^{\prime }}=0$ should be satisfied. In other words, the antiresonances correspond to the roots of $j_{n}(mx)=0$ or $[mxj_{n}(mx)]^{^{\prime }}=0$ exactly. In Fig. \ref{fig3}(b), we take the dipolar mode as an example. The first antiresonance at $x=0.501$ corresponds to $[mxj_{1}(mx)]^{^{\prime }}=0$, whereas the second antiresonance at $x=0.820$ corresponds to $j_{1}(mx)=0$. Although the corresponding scattering fields are zero outside the sphere at the antiresonances, the fields inside the sphere are not necessarily zero. For instance, by checking the fields inside the TI sphere at $x=0.501$, it is found that cross-polarized electric dipolar mode could be exited since $d_1^{TI}\neq0$ whereas cross-polarized magnetic dipolar mode does not exist since $c_1^{TI}=0$. On the contrary, for $x=0.820$ we have $c_1^{TI}\neq 0$ and $d_1^{TI}=0$ as shown in Fig. \ref{fig3}(d), which implies the existence of cross-polarized magnetic dipolar mode and the lack of cross-polarized electric dipolar mode.

\begin{figure}
\centering\includegraphics[width=10cm]{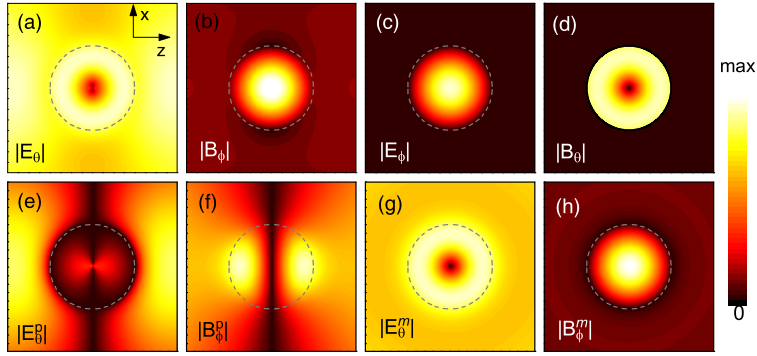}
\caption{Field distributions in scattering plane $\phi=0$ at antiresonances $x=0.501$. In (a) and (b), the field components $|E_\theta|$ and $|B_\phi|$ of total fields contributed by the superposition of electric and magnetic dipolar modes are shown respectively. In (c) and (d), the components $|E_\phi|$ and $|B_\theta|$ are trapped inside the sphere, which are contributed from the cross-polarized electric dipolar mode due to the TME effect. (e) and (f) show the field components $|E_\theta|$ and $|B_\phi|$ of the electric dipolar mode arised from bulk scattering; while in (g) and (h), $|E_\theta|$ and $|B_\phi|$ of the magnetic dipolar mode from bulk scattering are plotted. The field components $E_{\theta}^{p}$ and $B_{\phi}^{m}$ vanish at the boundary of the TI sphere. The physical parameters are the same as those in Fig. \ref{fig3}. The boundary of the TI sphere is shown by the dashed lines.}
\label{fig4}
\end{figure}

From the definition of the vector spherical harmonics \cite{boh83}, the tangential components (i.e., $\hat{\mathbf{e}}_\theta$ and $\hat{\mathbf{e}}_\phi$) of $\mathbf{N}_n$ and $\mathbf{M}_n$ are proportional to $[mxj_{n}(mx)]^{^{\prime }}$ and $j_{n}(mx)$ respectively. Thus, the underlying physics for the existence of anti-resonances of cross-polarized scattering coefficients is that the tangential field components corresponding to $\mathbf{N}_n$ or $\mathbf{M}_n$ vanish at the boundary of sphere. Note that the electric(magnetic) fields of electric multipoles are contributed from $\mathbf{N}_n$($\mathbf{M}_n$), while those of magnetic multipoles are proportional to $\mathbf{M}_n$($\mathbf{N}_n$). In other words, the corresponding tangential field components of the electric and magnetic multipoles should vanish at the boundary. In most cases, the electric and magnetic multipolar fields yield from the bulk scattering are strong and the cross-polarized ones induced by TME effect cannot manifest themselves. However, TME effect can be clearly observed at two particular scattering planes, i.e., planes with $\phi=0^\mathrm{o}$ and $\phi=90^\mathrm{o}$. In Fig. \ref{fig4}, we show the field distributions at scattering plane $\phi=0^\mathrm{o}$ for the first antiresonance at $x=0.501$ which corresponds to a root of $[mxj_{1}(mx)]^{^{\prime }}=0$. The upper panel of Fig. \ref{fig4} shows the total fields distribution, the dominant field components are $E_\theta$ and $B_\phi$ , which are the superposition of the fields from electric dipole (polarized in $x$ direction) and magnetic dipole (polarized in $y$ direction) modes due to the bulk scattering. The field patterns are almost the same with those for conventional dielectric spheres. Interestingly, the field components $E_\phi$ and $B_\theta$ induced by cross-polarized electric dipole(polarized along $y$ direction) are nonzero and trapped inside the TI sphere due to TME effect, whereas for conventional dielectric spheres these two components should be zero. As shown in Fig. \ref{fig4}(c) and (d), the cross-polarized fields induced by TME effect are completely trapped inside the sphere. Observation of the trapping effect can be made clearly since the incident and scattered waves don't have these two components($E_\phi$ and $B_\theta$) at the scattering plane $\phi=0^\mathrm{o}$. For conventional spheres \cite{sch:09, hsu14}, the scattered waves are reduced to zero at antiresonances and thus the spheres become transparent, however, the corresponding field components of incident waves still exist inevitably.

In contrast, at the antiresonances of TI cylinders, the electric multipolar modes (corresponding to $a_n$) and magnetic multiple modes (corresponding to $b_n$) due to the bulk scattering can be excited separately via waves with different polarization, i.e., TE waves and TM waves respectively. As is well known, for TI spheres electric multipolar modes ($a_n$) and magnetic multipolar modes ($b_n$) will always be excited simultaneously. Therefore, at the antiresonances, the total fields of bulking scattering at the boundary can not be reduced to zero in general. However, the tangential components of the electric and magnetic fields for respectively the electric and magnetic multipolar modes, are zero as shown in Fig. \ref{fig4}(e) and \ref{fig4}(h) for the dipolar case with $[mxj_{1}(mx)]^{^{\prime }}=0$ satisfied. The vanishment of the tangential field components at antiresonances \cite{ge14a} for electric and magnetic dipoles is an exotic phenomenon and also a strong evidence that the TME effect is intrinsically a surface effect. For the second antiresonance at $x=0.820$ similar conclusions can be drawn. The field trapping for the cross-polarized magnetic dipole (note that $c_n^{TI}\neq 0$ here) inside the sphere can also be observed.

\section{Determining the fine structure constant in Rayleigh limit}

A quantitative determination of the TME effect can also be found in TI spheres similar to the case in TI cylinder \cite{ge14b}.  We can conduct a Rayleigh-scattering experiment with incident plane waves as shown in Fig. \ref{fig1} schematically. In the far field, both electric-field components of scattered waves $|E_{s\parallel }|$ and $|E_{s\perp }|$ are measurable quantities and we can thus define a quantity $R(\theta ,\phi)=|E_{s\parallel }|/|E_{s\perp }|$. In the Rayleigh limit, the scattering is dominantly contributed from dipoles. Assuming that $\varepsilon _{b}=\mu_{b}=\mu _{1}=1$ we have:

\begin{equation}
R(\theta ,\phi )=\left\vert \frac{(\overline{\alpha }^{2}-b\cos \theta )\cos
\phi -3\overline{\alpha }(1-\cos \theta )\sin \phi }{(\overline{\alpha }%
^{2}\cos \theta -b)\sin \phi -3\overline{\alpha }(1-\cos \theta )\cos \phi }%
\right\vert
\label{alpha}
\end{equation}%
where $b=3m^{2}-3+2\overline{\alpha }^{2}$ is basically a bulk parameter. To extract the fine structure constant, we firstly focus on the scattering angles that're in the first quadrant, i.e. $\theta,\phi \in (0,\pi/2)$. Obviously, $\overline{\alpha }^{2}$ is a small term compared to $b$, $b\cos\theta$ and hence could be neglected except in the vicinity $\theta\sim \pi/2$. With these assumptions we have:
\begin{equation}
\overline{\alpha }= \frac{bR(\theta ,\phi) \sin \phi-b \cos\theta \cos \phi}{3(1-\cos \theta )[\sin \phi-R(\theta, \phi) \cos \phi]}
\end{equation}%
This offers a simple way to determine $\overline{\alpha }$ if $R(\theta ,
\phi)$ is measured and the parameter $m$ is known. Clearly, the parameter $m$ is material-dependent and so does this measurement. However, if we set a special observation anlge $(\theta,\phi)=(90^{\mathrm{o}},0^{\mathrm{o}})$, from Eq. (\ref{alpha}) we have:
\begin{equation}
\overline{\alpha }=3R(90^{\mathrm{o}},0^{\mathrm{o}})
\end{equation}%
Owing to the complete suppression of bulk scattering in this particular angle,
the to-be measured $\overline{\alpha }$ now can be solely determined by the value
of $R(90^{\mathrm{o}},0^{\mathrm{o}})$ which is material-independent. From
the obtained $\overline{\alpha }$, it offers a way to measure the fine
structure constant $\alpha$ since the axion angle is quantized.

\section{Conclusions}

We've studied EM wave scattering by TI spheres systematically. The scattering coefficients for TI spheres are derived analytically. We've found that the scattering patterns of TI spheres showing strong backward scattering with impedance-matched background $m\sim 1$. The enhanced backscattering or the cancellation of the forward scattering is non-trivial since it is a strong manifestation of the TME effect where the EM scattering due to the surface of TIs play a key role. The strong backward scattering has a broadband feature and holds for both Mie and Rayleigh scattering regimes due to the non-resonant nature of TME effect. The anti-resonances in the Mie scattering regime are another kind of novel phenomenon of TI spheres, wherein the cross-polarized fields induced by TME effect are trapped inside the TI spheres. Finally, we proposed a simple way to determine the quantized TME effect of TIs in the Rayleigh limit by measuring the electric-field components of scattered waves in the far field.

\section*{Acknowledgement}

This work is supported by the 973Program (Grant Nos. 2013CB632701 and 2011CB922004). The research of H.D.Z. is further supported by the National Natural Science Foundation of China (Grant No. 11304038) and the Fundamental Research Funds for the Central Universities (Grant No. CQDXWL-2014-Z005).


\end{document}